\documentclass[twocolumn, prb, aps, showpacs, letterpaper]{revtex4}

\usepackage[dvips]{graphicx} % Include figure files
\usepackage{subfigure}
\usepackage{latexsym}
\usepackage{dcolumn} % Align table columns on decimal point
\usepackage{bm} % bold math

\begin{document}
\title{
Pressure effects on charge, spin, and metal-insulator transitions in narrow bandwidth manganite Pr$_{1-x}$Ca$_{x}$MnO$_{3}$
}

\author{Congwu Cui}
\author{Trevor A. Tyson}
\affiliation{Physics Department, New Jersey Institute of Technology, Newark, New Jersey 07102}

\date{\today}

\begin{abstract}

Pressure effects on the charge and spin states and the relation between the ferromagnetic and metallic states were explored on the small bandwidth manganite Pr$_{1-x}$Ca$_{x}$MnO$_{3}$ (x = 0.25, 0.3, 0.35). Under pressure, the charge ordering state is suppressed and a ferromagnetic metallic state is induced in all three samples. The metal-insulator transition temperature (T$_{MI}$) increases with pressure below a critical point P*, above which T$_{MI}$ decreases and the material becomes insulating as at the ambient pressure. The e$_{g}$ electron bandwidth and/or band-filling mediate the pressure effects on the metal-insulator transition and the magnetic transition. In the small bandwidth and low doping concentration compound (x = 0.25), the T$_{MI}$ and Curie temperature (T$_{C}$) change with pressure in a reverse way and do not couple under pressure. In the x = 0.3 compound, the relation of T$_{MI}$ and T$_{C}$ shows a critical behavior: They are coupled in the range of $\sim$0.8-5 GPa and decoupled outside of this range. In the x = 0.35 compound, T$_{MI}$ and T$_{C}$ are coupled in the measured pressure range where a ferromagnetic state is present.

\end{abstract}

\pacs{71.27.+a, 71.30.+h, 75.25.+z, 75.47.Lx}
%62.50.+p High-pressure and shock wave effects in solids and liquids  
%75.47.-m Magnetotransport phenomena; materials for magnetotransport 
%71.27.+a Strongly correlated electron systems; heavy fermions  
%71.28.+d Narrow-band systems; intermediate-valence solids (for magnetic aspects, see 75.20.Hr and 75.30.Mb in magnetic properties and materials) 
%71.30.+h Metal-insulator transitions and other electronic transitions 
%75.25.+z Spin arrangements in magnetically ordered materials (including neutron and spin-polarized electron studies, synchrotron-source x-ray scattering, etc.)  
%75.47.Gk Colossal magnetoresistance  
%75.47.Lx Manganites

\maketitle

\section{Introduction}
In the manganite of Pr$_{1-x}$Ca$_{x}$MnO$_{3}$ (PCMO), the large size difference between the Pr/Ca and the Mn cations leads to a small tolerance factor and hence a small transfer integral between Mn atoms in the whole doping range.\cite{tokura_jap_79_5288_96} The e$_{g}$ electrons are localized and the charge ordering (CO) phase is stabilized in a large doping range. In the low doping range (0.15$\leq$x$<$0.3) and at low temperatures, Pr$_{1-x}$Ca$_{x}$MnO$_{3}$ is a ferromagnetic insulator. It is believed to exhibit an orbitally ordered ground state similar to that in LaMnO$_{3}$ and the cooperative Jahn-Teller distortion of MnO$_{6}$ octahedra leads to (3x$^{2}$ - r$^{2}$)/(3y$^{2}$ - r$^{2}$)-type orbital ordering in ab-plane, while the Mn$^{4+}$ ions are disordered and no charge ordering state is reported in this range.\cite{zimmermann_prb_64_195133_01} In the range of 0.3$\leq$x$\leq$0.7, the compounds are antiferromagnetic insulating (AFI) at low temperatures and the charges and orbitals are ordered. The charge ordered state is sensitive to external fields and radiations. Application of magnetic fields\cite{tomioka_jpsj_64_3626_95} and high electric fields,\cite{asamitsu_nature_388_50_97} irradiation by x-rays\cite{kiryukhin_nature_386_813_97, cox_prb_57_3305_98} or visible light\cite{miyano_prl_78_4257_97, mori_jpsj_66_3570_97} can all destroy the charge ordering and lead to a conducting state.

In compounds with a CO state, the lattice is strongly coupled to the spin and charge.\cite{dediu_prl_84_4489_00} While charges are ordered, local distortion changes from dynamic Jahn-Teller distortion to a collective static distortion\cite{dediu_prl_84_4489_00} and the MnO$_{6}$ octahedra buckle.\cite{yoshizawa_prb_52_R13145_95} Raman scattering in the x = 0.37 compound reveals that a strong coupling between the spin and lattice degrees of freedom induces the large variation of the mode frequency and anomalous line width broadening of the A$_{g}$(2) and A$_{g}$(4) phonons with temperature decrease.\cite{gupta_europl_58_778_02}

Under pressure, due to the bandwidth (W) increase, the CO state melts and a metallic state is induced. Moritomo \textit{et al}.\cite{moritomo_prb_55_7549_97} reported that pressure up to 0.8 GPa suppresses the CO of compound x = 0.35, 0.4, 0.5 and dT$_{CO}$/dP increases with x. In the x = 0.3 compound, pressure above 0.5 GPa induces a metallic transition which is assigned to a charge ordered insulating (COI) to ferromagnetic (FM) metallic (FMM) transition. Magnetic field was found to be almost equivalent to pressure up to 1.5 GPa and the effect of magnetic field can be scaled to pressure. The CO state is more robust when x is near to the commensurate value 0.5, so in the x = 0.35 compound, the insulator-metal transition was not found under pressures up to 1.6 GPa.

In many manganites with a metal-insulator transition, the metallic state and ferromagnetic state are coupled, which is explained by the double exchange theory. When a external pressure is applied, the Curie temperature (T$_{C}$) and the metal-insulator transition temperature (T$_{MI}$) still coincide in the low pressure range ($<$1.6 GPa).\cite{arnold_apl_67_2875_95,neumeier_prb_52_R7006_95} Because of the limit of high-pressure techniques in magnetic measurements, the question of whether the metallic state and ferromagnetic state are coupled at high-pressure is still open to date. It has been reported that the substitution of La atom with Gd and Y in the La-Ca-Mn-O manganites leads to a separation between T$_{C}$ and T$_{MI}$.\cite{terashita_prb_63_174436_01, mahendiran_prb_53_12160_96} The most important effect of the chemical substitution with the smaller Gd and Y ions is the bandwidth change. The decoupling of T$_{C}$ and T$_{MI}$ indicates that bandwidth plays a crucial role in the coupling of ferromagnetic and metallic states.

In this article, chemical doping and external pressure methods are combined to investigate the metal-insulator transition and the relation between the ferromagnetic and metallic states, in addition to the observation on charge ordering. With the application of pressure in the range from ambient to $\sim$6.5 GPa, the magnetic and electronic properties of the small bandwidth PCMO system (x = 0.25, 0.30, 0.35) have been found to be greatly affected. At high pressures, pressure effects are far more complicated than in low pressure range ($<\sim$2 GPa). In addition to the effects on the electronic states and charge ordering, the magnetic states are also modified by pressure. Under pressure, both the electronic transition and magnetic transition display critical behavior. The e$_{g}$ electron bandwidth and/or the band-filling mediate the behavior of the magnetic and electronic transitions under pressure. Only when the bandwidth and/or band-filling are large enough, can the magnetic and electronic transitions be coupled. For a critical bandwidth and/or band-filling, the relation between them also shows a critical behavior.

\section{Samples and Experimental Methods}
\begin{figure}
\subfigure{\label{fig1a} \includegraphics[width=2.5in]{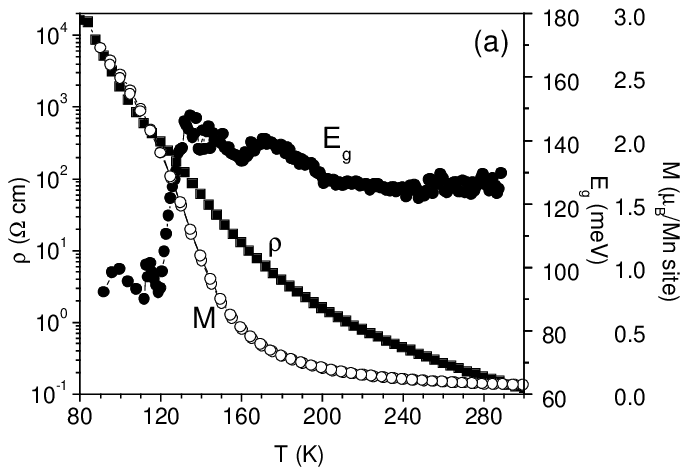}}\\
\subfigure{\label{fig1b}  \includegraphics[width=2.5in]{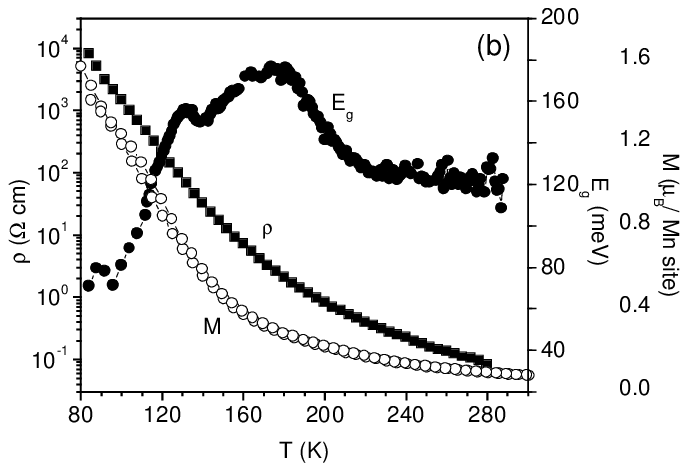}}\\
\subfigure{\label{fig1c}  \includegraphics[width=2.5in]{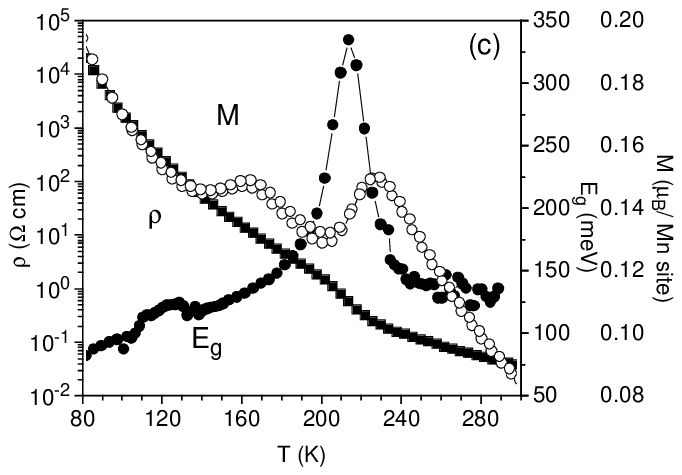}}\\
\caption{\label{fig_1}Resistivity, magnetization and activation energy of (a) Pr$_{0.75}$Ca$_{0.25}$MnO$_{3}$, (b) Pr$_{0.7}$Ca$_{0.3}$MnO$_{3}$, and (c) Pr$_{0.65}$Ca$_{0.35}$MnO$_{3}$ at ambient pressure. Magnetization (open circles) was measured from 4.2 K to 400 (field cooled and zero field cooled) at 10 kOe. Resistivity and the activation energy are represented by solid squares and solid circles respectively.}
\end{figure}

The samples were prepared by solid-state reaction. The procedure and details of making the samples were described elsewhere.\cite{cui_apl_83_2856_03} The x-ray diffraction patterns taken at room temperature show that the samples are in a single crystallographic phase. The structure was refined to Pbnm symmetry using the Rietveld method. All samples have the O$'$-type orthorhombic structure ($b>a>\frac{c}{\sqrt{2}}$). The samples were characterized by magnetization measurements [Figs.\ \ref{fig1a}, \ref{fig1b}, and \ref{fig1c}]. The results agree with those published by other groups.\cite{dediu_prl_84_4489_00, yoshizawa_prb_52_R13145_95, tokura_jmmm_200_1_99} The details of the high-pressure resistivity measurement method and error analysis were described previously.\cite{cui_prb_67_104107_03}

It has been extensively reported that the electrical transport of the PCMO system is highly nonlinear at low temperature.\cite{asamitsu_nature_388_50_97, mercone_prb_65_214428_02, guha_prb_62_R11941_00, guha_prb_62_5320_00, stankiewicz_prb_62_11236_00} Because this paper is focused on the pressure effects, to avoid the complexity induced by the electrical field, a low constant bias current in the Van der Pauw method was used to maintain the transport behavior in the ohmic regime as the instrument noise allows. The size of the samples used for high-pressure resistivity measurements is $\sim$$250\times250\  \mu m^{2}$. For the PCMO compounds, the bias current of 10 $\mu A$  gives a electrical field in the order of $\sim$10$^{2}$ V/cm at liquid nitrogen temperature, similar to the ohmic case in Ref.\ \onlinecite{asamitsu_nature_388_50_97} and far below the nonlinear region probed in Ref.\ \onlinecite{asamitsu_nature_388_50_97, mercone_prb_65_214428_02, guha_prb_62_R11941_00, guha_prb_62_5320_00, stankiewicz_prb_62_11236_00}. However, a recent report\cite{markovich_prb_68_094428_03} indicates that the nonlinear behavior can persist even at lower electrical field, so that the resistivity and activation energy below the ferromagnetic transition are temperature dependent. In this article, the equivalent electrical field due the bias current is the same order as in Ref.\ \onlinecite{markovich_prb_68_094428_03}. Although the measured I-V curves at temperatures below the magnetic transition are not completely ohmic in the measured temperature range, the applied current is still well below the negative resistive region. Moreover, the I-V curves become more ohmic under pressure due to the material becoming more metallic. Hence, the pressure dependent trend is not affected.

In PCMO system, from x = 0.1 to 0.4, the resistivity displays p-type semiconducting behavior, $\rho=\rho_{0}\text{exp}(E_{g}/k_{B}T)$, with the activation energies (E$_{g}$) being slightly above 100 meV at room temperature.\cite{jirak_jmmm_53_153_85} The activation energy can be determined numerically by calculating $dln(\rho)/d(k_{B}T)^{-1}$ from resistivity data. Figure \ref{fig_1} gives the resistivity, magnetization, and activation energy E$_{g}$ of the samples at ambient pressure as a function of temperature. The E$_{g}$ of the samples is  $\sim$125 meV and does not change with temperature in the paramagnetic phase near room temperature, showing semiconducting behavior, as reported by Jir\'ak \textit{et al}.\cite{jirak_jmmm_53_153_85}

On the E$_{g}$ curves, upon cooling, there is an increase at $\sim$200 K, $\sim$220 K, $\sim$240 K for each sample, respectively. Apparently, the increase in x = 0.3 and 0.35 corresponds to the charge ordering. For the x = 0.25 sample, the charge ordering and orbital ordering state was predicted theoretically,\cite{mizokawa_prb_63_024403_00, takashi_prb_61_11879_00} but it  was not observed by the x-ray resonant scattering studies.\cite{zimmermann_prb_64_195133_01} In our transport measurements, compared with the other two samples (x = 0.30, 0.35) with charge ordering state, the E$_{g}$ arise at $\sim$200 K seems to correlate to the charge ordering transition. For the x = 0.35 sample, corresponding to a sharp peak at $\sim$215 K is the charge ordering state, as in a similar compound.\cite{rivadulla_solidss_110_179_99} The peaks corresponding to charge ordering have different heights ($\sim$350 meV in x = 0.35, $\sim$170 meV in x = 0.3, and $\sim$140 meV in x = 0.25), indicating that the charge ordering is the most robust in x = 0.35 and the weakest in x = 0.25.

In the x = 0.25 and x = 0.3 samples, there is a sharp drop of E$_{g}$ on cooling in the range of $\sim$100-140 K. The x = 0.25 compound is ferromagnetic insulating at low temperatures, while the x = 0.3 compound is ferromagnetic and antiferromagnetic phase-separated.\cite{jirak_jmmm_53_153_85, frontera_prb_62_3381_00} Below the FM transition, the material is still insulating but with a much smaller energy gap than in paramagnetic phase, indicating that the ferromagnetically coupled spins enhance the electron transport. This  is correlated with the magnetic semiconductor behavior, corresponding to a orthorhombic strain change.\cite{markovich_prb_68_094428_03} By comparing the temperature dependence of resistivity, magnetization, and E$_{g}$ [Figs.\  \ref{fig1a} and \ref{fig1b}], the drastic reduction of E$_{g}$ can be correlated with the ferromagnetic transition.\cite{cui_apl_83_2856_03} The T$_{C}$ of the x = 0.25 compound at ambient pressure determined from the activation energy as a function of temperature is 124$\pm$4 K, consistent to that determined by magnetization measurements by other authors.\cite{fisher_prb_68_174403_03} For the x = 0.35 sample, in the temperature range of $\sim$100-140 K on the E$_{g}$ curve, there is a small bump, which may correspond to the CE-type AFI transition [Fig.\ \ref{fig1c}]. In the measured temperature range, the low temperature magnetic state is antiferromagnetic, displaying in E$_{g}$ as a slow change with temperature.

Due to the changes in the activation energy with temperature during the magnetic and charge ordering transitions, the high-pressure effects on charge and spin states can be observed simultaneously through resistivity measurement. In this article, the resistivity measurements have been performed more times than displayed at different pressures for each sample, but only the resistivity at several typical pressures are shown to avoid confusion. The pressure induced metal-insulator transition temperature is defined as the peak temperature of the resistivity. The magnetic transition temperature is defined as the point at which the activation energy changes fastest by referring the first derivative. For the charge ordering transition, with the present data, T$_{CO}$ cannot be extracted accurately, so the charge state change under pressure in the x = 0.25 and 0.3 samples are only qualitatively described. For the x = 0.35 sample, the E$_{g}$ peak is defined as T$_{CO}$. In this way, although T$_{CO}$ is lower than the temperature at which the charge ordering actually starts to appear on cooling, it still describes the charge state changes under pressure correctly.

\section{Results and Discussions}

\subsection{Pr$_{0.75}$Ca$_{0.25}$MnO$_{3}$}

\begin{figure}
\subfigure{\label{fig2a} \includegraphics[width=2.25in]{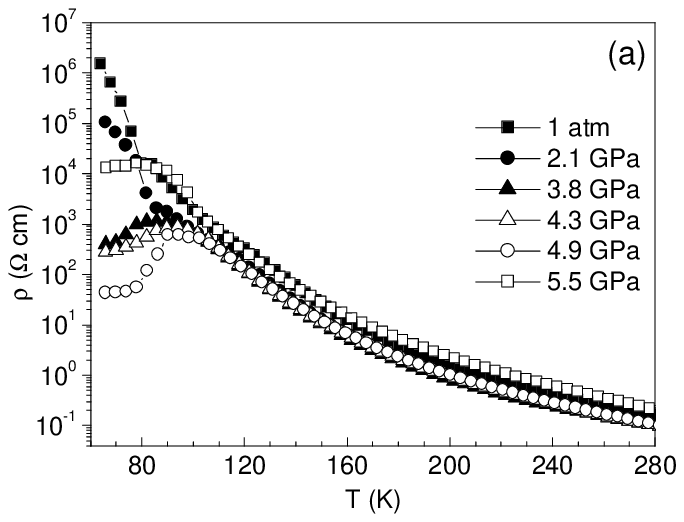}}\\
\subfigure{\label{fig2b}  \includegraphics[width=2.25in]{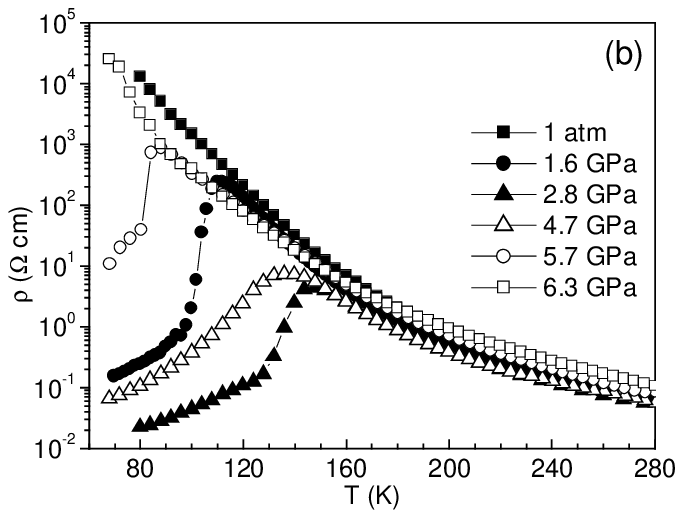}}\\
\subfigure{\label{fig2c}  \includegraphics[width=2.25in]{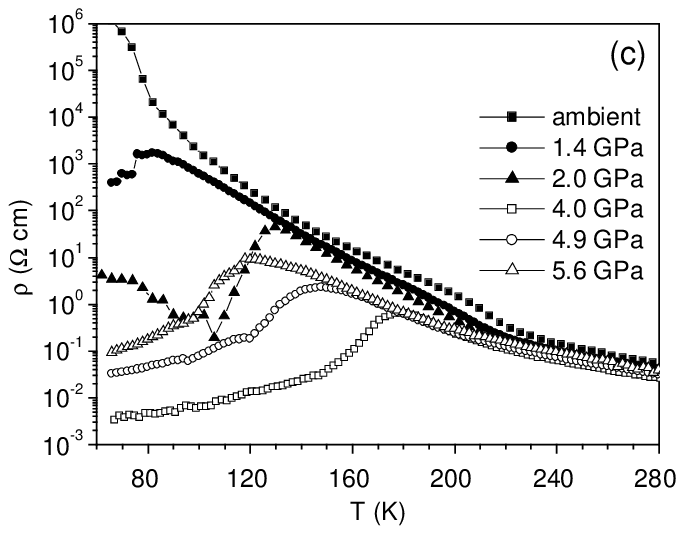}}\\
\caption{\label{fig_2}Resistivity of (a) Pr$_{0.75}$Ca$_{0.25}$MnO$_{3}$, (b) Pr$_{0.7}$Ca$_{0.3}$MnO$_{3}$, and (c) Pr$_{0.65}$Ca$_{0.35}$MnO$_{3}$ as a function of temperature under pressure.}
\end{figure}
\begin{figure}
\subfigure{\label{fig3a} \includegraphics[width=2.5in]{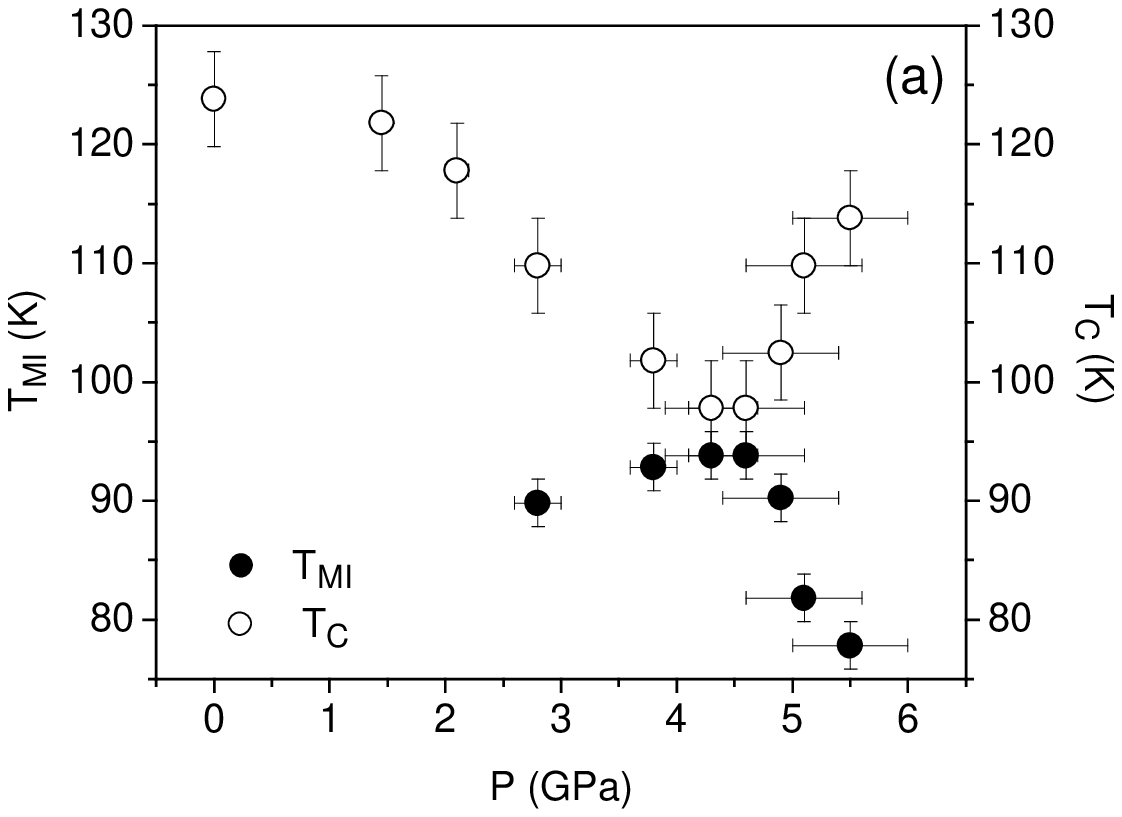}}\\
\subfigure{\label{fig3b}  \includegraphics[width=2.5in]{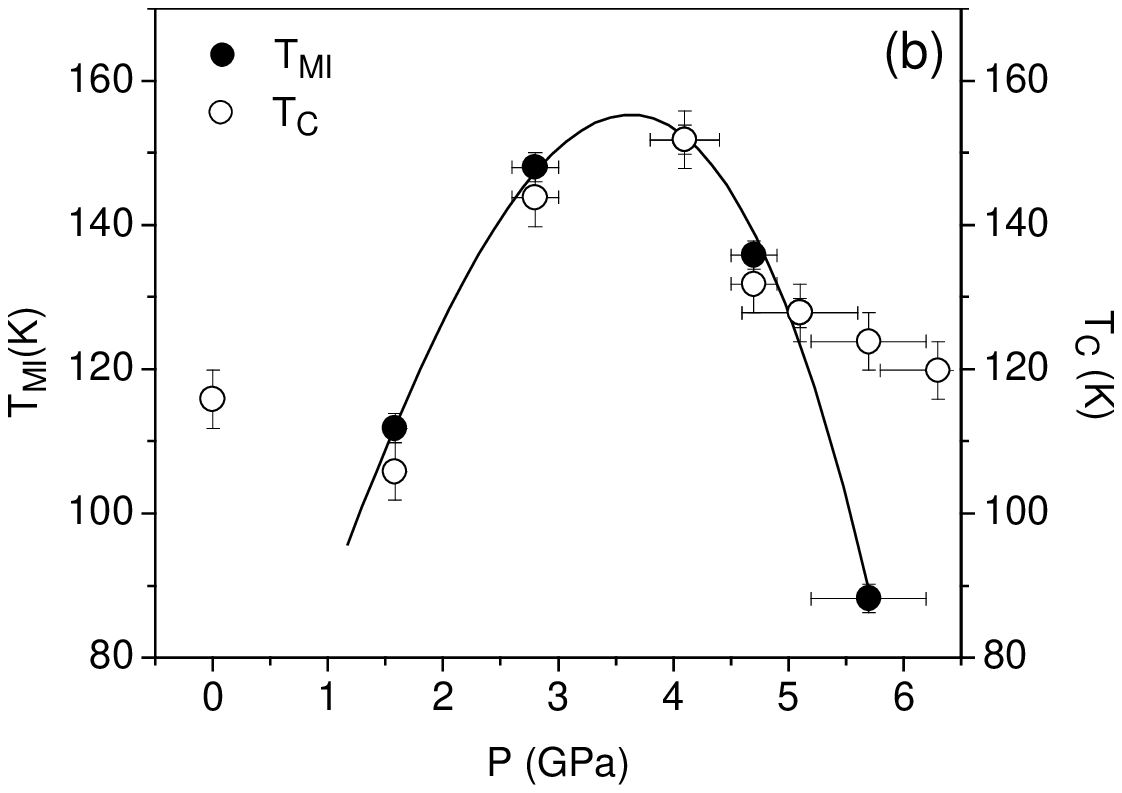}}\\
\subfigure{\label{fig3c}  \includegraphics[width=2.5in]{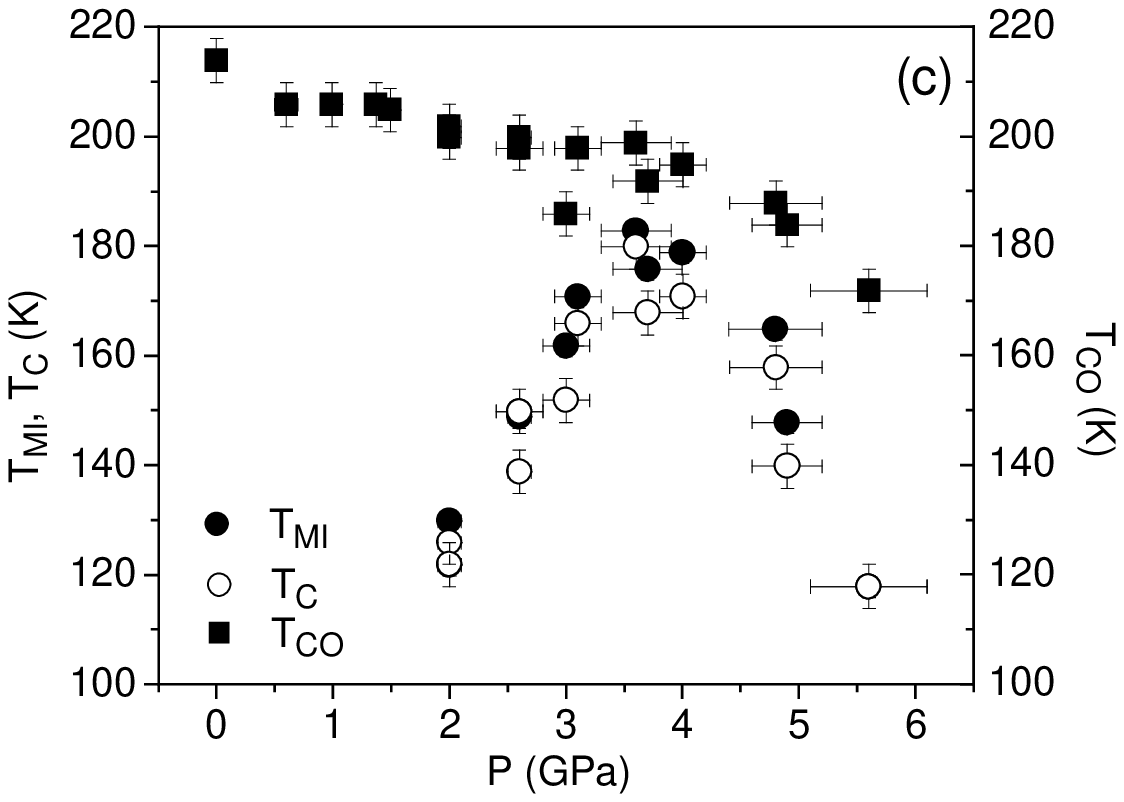}}\\
\caption{\label{fig_3}Transition temperatures of (a) Pr$_{0.75}$Ca$_{0.25}$MnO$_{3}$, (b) Pr$_{0.7}$Ca$_{0.3}$MnO$_{3}$, and (c) Pr$_{0.65}$Ca$_{0.35}$MnO$_{3}$. Note: the solid circles represent T$_{MI}$; the open circles represent T$_{C}$ extracted from the activation energy; the solid squares in (c) represent the charge ordering temperature; the solid line in (b) is a fit to T$_{MI}$ with a third-order polynomial as a guide to the eye.}
\end{figure}

In Fig.\ \ref{fig2a} is the resistivity of Pr$_{0.75}$Ca$_{0.25}$MnO$_{3}$ at several pressures. With pressure increase, the low temperature insulating state is suppressed. Although there is a trace of transition, the resistivity only shows a shoulder below  $\sim$3 GPa, above which the resistivity peak corresponding to the metal-insulator transition starts to become evident. The transition temperature increases with pressure. When the pressure is higher than a certain value, the trend reverses so that the low temperature state does not become more metallic, but insulating, with the transition temperature decreasing simultaneously. The metal-insulator transition temperature is shown as a function of pressure in Fig.\ \ref{fig3a}. The resistivity in the paramagnetic and ferromagnetic phases follows the same trend as the transition temperature.

\begin{figure}
\subfigure{\label{fig4a} \includegraphics[width=2.25in]{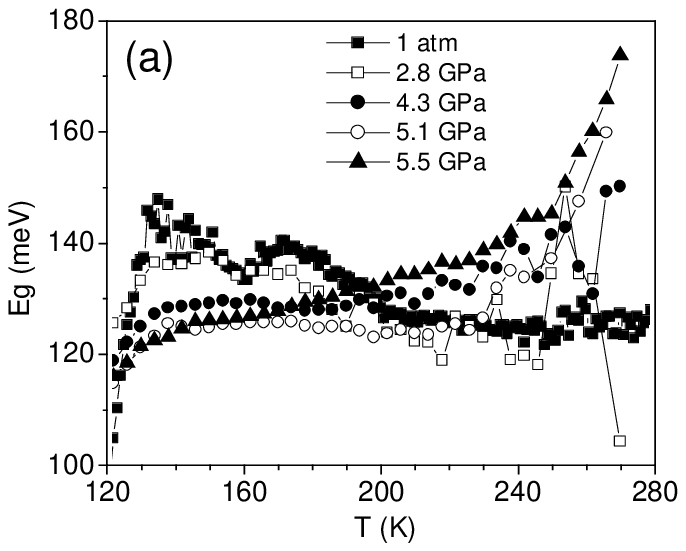}}\\
\subfigure{\label{fig4b}  \includegraphics[width=2.25in]{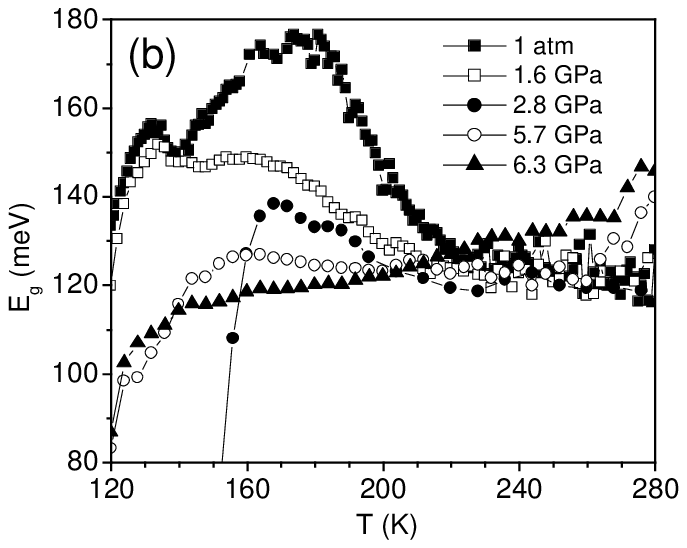}}\\
\subfigure{\label{fig4c}  \includegraphics[width=2.25in]{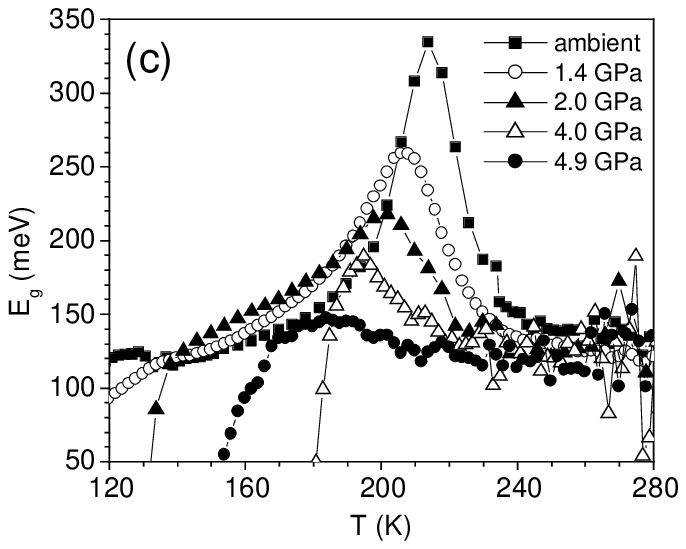}}\\
\caption{\label{fig_4}Activation energy of (a) Pr$_{0.75}$Ca$_{0.25}$MnO$_{3}$, (b) Pr$_{0.7}$Ca$_{0.3}$MnO$_{3}$, and (c) Pr$_{0.65}$Ca$_{0.35}$MnO$_{3}$ under pressure.}
\end{figure}
E$_{g}$ \textit{vs}.\  T plots are shown in Fig.\ \ref{fig4a}. With pressure increase, E$_{g}$ increase upon cooling is suppressed, indicating that the charge ordering state is suppressed. While the CO state is suppressed, E$_{g}$ becomes temperature dependent in the paramagnetic phase in high-pressure range so that E$_{g}$ increases on warming up and the temperature dependency of E$_{g}$ becomes stronger with pressure increase. The behavior of the electron transport in the paramagnetic phase at high pressures cannot be fit to the known polaron hoping and variable range hoping models.\cite{cui_apl_83_2856_03} The origin is still not understood.

In Fig.\ \ref{fig4a}, corresponding to the ferromagnetic transition is the reduction of the activation energy. The transition temperatures extracted are displayed in Fig.\ \ref{fig3a} together with T$_{MI}$. In this compound, the magnetic transition and electronic transition are significantly decoupled. Below $\sim$4 GPa, with pressure increase, the ferromagnetic state is suppressed, displaying as T$_{C}$ decreases while the electronic transition temperature T$_{MI}$ increases. This indicates that the conducting mechanism in the low temperature range is not double exchange and that a competing mechanism takes effect. In a lower pressure range ($<$1.2 GPa), Markovich et al.\ \cite{markovich_prb_68_094428_03} reported that T$_{C}$ of Pr$_{0.8}$Ca$_{0.2}$MnO$_{3}$ increases linearly with pressure from $\sim$130 K at ambient pressure to $\sim$132 K at $\sim$1.2 GPa. In Fig.\ \ref{fig3a}, the behavior of T$_{C}$ of the x=0.25 compound below 1.2 GPa can not be determined due to the large error bars. However, it can be clearly seen that the rate of change of T$_{C}$ with pressure is much slower below 1.2 GPa than that above 1.2 GPa

In thin films, it was found that the metal-insulator transition and the magnetic transition are decoupled,\cite{aarts_apl_72_2975_98, rao_jap_85_4794_99} which may be ascribed to the strong disorder at T$_{C}$. The disorder is overcome by magnetization increase while cooling, inducing a metallic state.\cite{aarts_apl_72_2975_98} The T$_{MI}$ and T$_{C}$ decoupling also exists in bulk materials Pr$_{0.7}$Ba$_{0.3}$MnO$_{3}$,\cite{heilman_prb_65_214423_02} in which it is ascribed to the competition between the double exchange and the superexchange interactions between neighboring Mn-Mn spins. The strength of superexchange is a function of bandwidth W. In low pressure range, because of the large lattice distortion, W is small, superexchange may dominate and hence, the material is insulating. With pressure increase, due to the local distortion suppression (indicated by the charge ordering disappearance), W increases, and the insulating state is suppressed correspondingly. The superexchange between two neighboring Mn$^{3+}$ can either be ferromagnetic or antiferromagnetic depending on the Mn-Mn distance.\cite{goodenough_pr_100_564_55} With pressure increase, the Mn-Mn distance monotonically decreases, the ferromagnetic superexchange interaction between the Mn$^{3+}$  cations is weakened so that T$_{C}$ decreases. On one hand, pressure suppresses the local distortion to enhance the metallic state; on the other hand, the decrease of Mn-Mn distance leads to T$_{C}$ decreasing. However, this is only the case below $\sim$4 GPa, above this pressure, T$_{C}$ increases with pressure. When pressure is above $\sim$4.5 GPa, the trends of both the magnetic and electronic transition temperature changes with pressure are reversed, indicating that the strength of the superexchange is a function of not only the Mn-Mn distance, but also possibly the structure (especially the local atomic structure of the MnO$_{6}$ octahedra).

\subsection{Pr$_{0.7}$Ca$_{0.3}$MnO$_{3}$}

In Fig.\ \ref{fig2b}, the temperature dependence of resistivity of Pr$_{0.7}$Ca$_{0.3}$MnO$_{3}$ at several pressures is shown. At ambient pressure, the material is insulating in the whole temperature range. As reported, at a pressure above 0.5 GPa, an insulating to metallic transition is induced, which is ascribed to a COI to FMM transition.\cite{moritomo_prb_55_7549_97, hwang_prl_75_914_95} With pressure increase, the transition temperature T$_{MI}$ is shifted to higher temperature and resistivity is suppressed. In the pressure range $\sim$3-4 GPa, the trend saturates. At higher pressure, T$_{MI}$ decreases and the resistivity increases. At $\sim$6.3 GPa, the material becomes insulating in the measured temperature range and the resistivity as a function of temperature almost reproduces the case at ambient pressure. T$_{MI}$ \textit{vs}.\  pressure is plotted in Fig.\ \ref{fig3b}.

E$_{g}$ as a function of temperature at different pressures are plotted in Fig.\ \ref{fig4b}. At ambient pressure, above $\sim$220 K, E$_{g}$ is $\sim$125 meV, then increases on cooling, implying the charge ordering. The E$_{g}$ arising on cooling disappears gradually with pressure increase and at $\sim$5.7 GPa, the CO state is completely suppressed so that the activation energy does not change with temperature and therefore, the material displays a semiconducting behavior above the magnetic transition. Same as in x = 0.25 sample, at high pressures, the activation energy becomes temperature dependent near to room temperature in the paramagnetic phase and increases upon warming up.

The magnetic transition temperatures extracted from the E$_{g}$ curves are shown in Fig.\ \ref{fig3b} together with T$_{MI}$. It is clearly seen that in the measured pressure range of $\sim$1.5-5GPa, T$_{C}$ and T$_{MI}$ coincide, suggesting that pressure destroys the charge ordering insulating state completely and induces a FMM state at low temperature. But in the low pressure range and above $\sim$5 GPa, the magnetic transition and metal-insulator transition are decoupled and the material becomes insulating at pressures near to ambient pressure and above $\sim$5 GPa.

In the medium pressure range, at optimum pressure, both the magnetic transition and metal-insulator transition temperature reach a maximum. This behavior is similar to that observed in the manganites with a larger bandwidth, in which it can be ascribed to the pressure induced Jahn-Teller distortion and Mn-O-Mn bond angle changes according to the double exchange theory.\cite{cui_prb_67_104107_03, congeduti_prl_86_1251_01, meneghini_prb_65_012111_02}

In the low ($<\sim$0.8 GPa) and high ($>\sim$5 GPa) pressure range, the material is more insulating and T$_{MI}$ and T$_{C}$ are decoupled. Neutron diffraction suggests that this compound could be considered as FM and AFM phase-separated at low temperature.\cite{jirak_jmmm_53_153_85} The decoupling behavior may be similar to that in the x = 0.25 compound. The difference is that Pr$_{0.75}$Ca$_{0.25}$MnO$_{3}$ is more distorted and has a smaller bandwidth so that under high-pressure T$_{MI}$ and T$_{C}$ never meet each other.

The charge ordering phase in the PCMO system is correlated with the lattice distortion and the buckling of the MnO$_{6}$ octahedra.\cite{yoshizawa_prb_52_R13145_95} At T$_{CO}$, a transition from dynamic Jahn-Teller distortion to collective static distortion takes place.\cite{dediu_prl_84_4489_00} Therefore, the CO state disappearance under pressure indicates that the octahedra buckling and the static Jahn-Teller distortions are suppressed and only the dynamic distortion is present at pressure above the optimum pressure.

In the electronic and magnetic transition temperatures \textit{vs}.\  pressure phase diagram [Fig.\ \ref{fig3b}], in the high ($>$$\sim$5 GPa) and low ($<$$\sim$0.8 GPa) pressure range, the magnetic and electronic phase transitions are decoupled. The compound is insulating at the ambient and high-pressures. The decoupling of T$_{C}$ and T$_{MI}$ has some similarities to the x = 0.25 compound. It is possible that pressure induces a ferromagnetic insulating state in the high-pressure range.\cite{cui_apl_83_2856_03}

\subsection{Pr$_{0.65}$Ca$_{0.35}$MnO$_{3}$}

\begin{figure}
\includegraphics[width=3in]{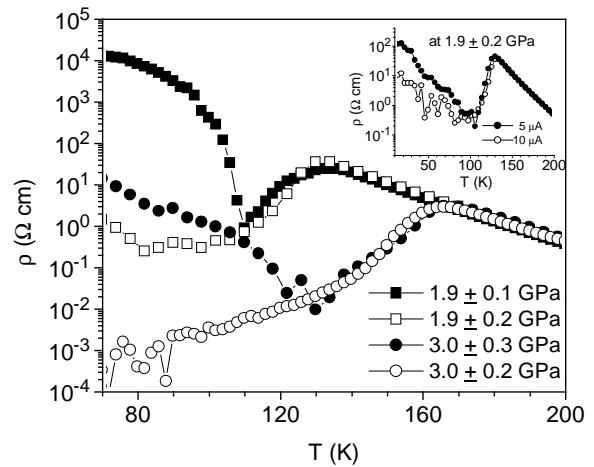}\\
\caption{\label{fig_5}Illustration to the joint effects of electrical field and pressure in Pr$_{0.65}$Ca$_{0.35}$MnO$_{3}$. The resistivity data come from four samples and at two pressures (see text for details). The inset shows the resistivity of one sample at $\sim$2 GPa and at different bias source currents.}
\end{figure}

Figure \ref{fig2c} shows the resistivity of the x = 0.35 sample under pressure. In the low pressure region, it appears that a metallic state is induced by pressure in the medium temperature range, and the T$_{MI}$ increases on pressure increase, while the low temperature state is still insulating. However, it is found that the low temperature insulating state is unstable in different loadings of sample. At the same pressure, the low temperature state can be either insulating or metallic from sample to sample. Figure \ref{fig_5} shows the resistivity of four different loadings of sample at two pressures with the same bias current. In these four samples, the low temperature state is much different even under almost the same pressures. One might guess that the difference in the low temperature state comes from the non-uniformity of the polycrystal sample due to the very small size. The four samples were cut from the same piece of size $\sim$$2\times 2 \  mm^{2}$ which came from a pellet. In addition, the difference occurs only in the low temperature range. In the high temperature range, the resistivity of different samples at the same pressure completely overlap. Considering the nonlinear behavior of the electrical transport in the low temperature range, the difference between samples at same pressure can be ascribed to the electrical field effect. The electrical contacts used for the Van der Pauw resistivity measurement were made by attaching four gold wires to the corners of the sample with silver paste. In this way, the distance between the leads cannot be controlled precisely due to the tiny sample size, thus leading to the variable electrical field from sample to sample. This is proved by the measurements on one of the samples in Fig.\ \ref{fig_5} at $\sim$2 GPa and at different bias current (Inset of Fig.\ \ref{fig_5}). When the bias current is decreased from 10 $\mu A$ to 5 $\mu A$, the resistivity in the low temperature range is increased. Therefore, electrical field and pressure jointly affect the low temperature state. At ambient pressure, the low temperature state is associated with the spin canted antiferromagnetic insulating state (CAFI).\cite{tokura_jmmm_200_1_99} So under pressure, both the CE-type AFI state and the CAFI state are suppressed. Unfortunately, Because of the electrical field effect, it is impossible to quantify the pressure effect on the CAFI state in our experiment.

Different from the CAFI state, the metallic state and the paramagnetic state are not affected by the electrical field. With pressure increase, the low temperature insulating state is completely suppressed and the material becomes ferromagnetic metallic. This can be seen in the E$_{g}$ \textit{vs}.\  temperature curves [Fig.\ \ref{fig4c}]. At high pressures at which the CAFI state is completely suppressed, the E$_{g}$ as a function of temperature is the same as the other two samples with ferromagnetic phases. T$_{MI}$ and T$_{C}$ extracted from E$_{g}$ \textit{vs}.\  pressure are plotted in Fig.\ \ref{fig3c}. Apparently, T$_{C}$ is coupled to T$_{MI}$ and follows the same manner as T$_{MI}$ in the measured pressure range. Similar to the other two samples, there is also a critical pressure for the metal-insulator transition temperature.

The pressure effect on the charge ordering is more evident in this sample. With pressure increase, charge ordering is suppressed and the transition temperature is shifted to low temperature and eventually the charge ordering transition disappears [Fig.\ \ref{fig4c}]. The charge ordering transition temperature as a function of pressure is plotted together with T$_{MI}$ and T$_{C}$ in Fig.\ \ref{fig3c}. Because the charge ordering transition temperature, corresponding to the E$_{g}$ increases upon cooling, is hard to define in the E$_{g}$ plots, the peak temperature of E$_{g}$ is used to represent the charge ordering transition. In this way, T$_{CO}$ is lower than what it appears to be. From the E$_{g}$ \textit{vs}. temperature curves, it cannot be determined if E$_{g}$ is temperature dependent in paramagnetic phase in high-pressure range. Same as in the x = 0.30 compound, the disappearance of the charge ordering also implies the suppression of the static Jahn-Teller distortion.

\subsection{External pressure  via.\ chemical doping}

In the PCMO system, the small size of the Pr$^{3+}$ ion induces larger lattice disorder than in the La-Ca-Mn-O system. The size difference between the Pr$^{3+}$ ion and the Ca$^{2+}$ ion is small (Ca$^{2+}$ ion is only slightly bigger than Pr$^{3+}$ ion).\cite{shannon_aca_32_785_76} Therefore, the e$_{g}$ electron bandwidth increase due to x increase is not expected to be large. Hence, the system is insulating in the whole doping range at ambient pressure. However, when external pressure is applied, the maximum pressure-induced metal-insulator transition temperature T$_{MI}$ at the critical pressure P* is much different [$\sim$90 K for x = 0.25; $\sim$155 K for x = 0.3, and $\sim$180 K for x = 0.35 (Fig.\ \ref{fig_3})]. This suggests that the bandwidth difference between the three samples may not be the only factor determining the pressure effects on the metal-insulator transition. On the other hand, the band-filling or the electrons in the e$_{g}$ band may also mediate the pressure effect in this system. Actually, the difference between the ground states in the x = 0.25, 0.30, 0.35 compounds has already shown the band-filling effects. When x is reduced from 0.5, the spin arrangement along the c-axis in the Pbnm symmetry changes from anti-parallel to parallel so that at x=0.3, the spin coupling along the c-axis is almost FM with the CE-type ordering being maintained in the ab-plane.\cite{jirak_jmmm_53_153_85,yoshizawa_prb_52_R13145_95}  The extra electrons on the Mn$^{4+}$ sites with a concentration of (0.5-x)/Mn site mediate the double exchange interaction along the c-axis.\cite{jirak_jmmm_53_153_85} The effect of the extra electrons on the Mn$^{4+}$ sites on the magnetic structure is enhanced with decrease in x from 0.5. Thus, the minimum magnetic field destroying the CO state decreases with x decrease.\cite{tomioka_prb_53_R1689_96} However, when the CO state is melted and well above the critical magnetic field, the sample with higher x has a higher metal-insulator transition temperature.\cite{tomioka_prb_53_R1689_96} This is analogous to the pressure effect. Therefore, the pressure effect on the metal insulator transition in the PCMO system is also critically tuned by the band-filling.

The coupling of the magnetic and electronic transition under pressure is an interesting topic in this system. Because of the small bandwidth, when superexchange dominates, the material is insulating and T$_{MI}$ and T$_{C}$ may be decoupled due to the competition between double exchange and superexchange interactions.

In the x = 0.25 compound, superexchange is so strong that even under pressure T$_{MI}$ and T$_{C}$ never coincide [Fig.\ \ref{fig3a}]. With pressure increase, T$_{C}$ decreases and T$_{MI}$ increases, indicating that superexchange is suppressed due to the bandwidth increase. Above the critical pressure, relative to the bandwidth, the ferromagnetic superexchange between Mn sites may dominate again and therefore, T$_{C}$ increases and T$_{MI}$ decreases. The x = 0.30 doping is a critical point. Under pressure, the relation between T$_{MI}$ and T$_{C}$ in this compound displays critical behavior: at low pressure near to ambient and high-pressure above $\sim$5 GPa where there are larger distortions and hence, smaller bandwidth, the compound is insulating and T$_{MI}$ and T$_{C}$ are decoupled. At the medium pressure, the bandwidth is large enough, superexchange is suppressed, and the ferromagnetic state is coupled to the metallic state [Fig.\ \ref{fig3b}]. For the x = 0.35 sample, T$_{MI}$ and T$_{C}$ are coupled. Therefore, it can be concluded that bandwidth plays a crucial role in determining if the ferromagnetic and metallic state can be coupled under pressure. In addition, as discussed above, due to the effect of the band-filling on the double exchange interaction along the c-axis, probably the doping concentration x is also important to determine if the T$_{C}$ and T$_{MI}$ can couple.  Moreover, it is also shown that no matter what bandwidth the sample has, pressure cannot always increase the transition temperatures, but there is a critical pressure above which the trend is reversed.

Charge ordering is another interesting feature in the PCMO system. The charge ordering state in x = 0.35 sample is the strongest one in the three samples. Under pressure, charge ordering states in all three samples are suppressed. In the high-pressure range, for the x = 0.25 and 0.30 compound, an insulating state with unknown conducting mechanism appears in paramagnetic phase. In this state, E$_{g}$ increases on warming. It is speculated that dynamic Jahn-Teller distortion exists in this phase.

\section{Summary}

To summarize, pressure can destroy the low temperature charge ordered insulating state and induce a metallic state in the PCMO system by increasing the bandwidth. Both the bandwidth and the band-filling may affect the pressure effects on the properties critically. The bandwidth competes with superexchange between the Mn sites, determining together with the band-filling if there is a possibility for the metal-insulator transition and magnetic transition to couple under pressure. When the ground state bandwidth and/or the band-filling is small, even under high pressure, the magnetic and electronic transition cannot be coupled. Compounds with larger bandwidth and/or band-filling have higher T$_{MI}$ at P*. In all samples, the charge ordering states are suppressed, suggesting the suppression of the static Jahn-Teller distortions. Due to the strong interactions between charge, spin, and lattice, high-pressure structural measurements, especially at multiple temperatures, are desired to explain the pressure effects on the magnetic and electronic states.

\begin{acknowledgments}
This work was supported by National Science Foundation under grant No.\ DMR-9733862 and DMR-0209243.
\end{acknowledgments}

%\bibliography{c:/cuicw/papers/bib_cuicw}
\bibliography{bib_cuicw}

\end{document}